\documentclass{jpp}
\usepackage{graphicx}

\usepackage[utf8]{inputenc}
\usepackage[T1]{fontenc}
\usepackage{amsmath}
\usepackage[hyperfootnotes=true]{hyperref} 

\shorttitle{Applying machine learning methods to laser acceleration of protons}                                   
\shortauthor{Desai et al.}

\title{Applying machine learning methods to laser
acceleration of protons: lessons learned
from synthetic data}

\author{Ronak Desai \aff{1} \corresp{\email{desai.458@osu.edu}},
        Thomas Zhang \aff{1},
        John J. Felice \aff{1},
        Ricky Oropeza \aff{1},
        Joseph R. Smith \aff{2},
        Alona Kryshchenko \aff{3},
        Chris Orban \aff{1},
        Michael L. Dexter \aff{4} and
        Anil K. Patnaik \aff{4}
}

\affiliation{
\aff{1}Department of Physics, The Ohio State University, Columbus, OH 43210, USA
\aff{2}Department of Physics, Marietta College, Marietta, OH 43750, USA
\aff{3}Department of Mathematics, California State University Channel Islands, Camarillo, CA 93012, USA
\aff{4}Air Force Institute of Technology, Wright-Patterson AFB, OH 45433, USA}

\begin{document}

\maketitle

\begin{abstract}
In this study we consider three different machine learning methods -- a two-hidden layer neural network, Support Vector Regression and Gaussian Process Regression -- and compare how well they can learn from a synthetic data set for proton acceleration in the Target Normal Sheath Acceleration regime. The synthetic data set was generated from a previously published theoretical model by Fuchs et al. 2005 that we modified. Once trained, these machine learning methods can assist with efforts to maximize the peak proton energy, or with the more general problem of configuring the laser system to produce a proton energy spectrum with desired characteristics. In our study we focus on both the accuracy of the machine learning methods and the performance on one GPU including the memory consumption.  Although it is arguably the least sophisticated machine learning model we considered, Support Vector Regression performed very well in our tests. 
\end{abstract}

\section{Introduction}
  \label{sec:intro}

The field of ultra-intense laser science is increasingly beginning to embrace machine learning methods \citep{Anirudh_etal2022,dopp_etal2023}. This is especially true as the repetition rates of ultra-intense laser systems increase and data acquisition systems improve (e.g. \citet{Heuer_etal2022}). However, to date, there has been limited use of machine learning (ML) to enhance and control proton acceleration from ultra-intense laser systems.  Recently, \citet{Loughran_etal2023} provide results from training a ML model on proton acceleration data using a laser system that operates at 1~Hz repetition rate using Bayesian optimization. \citet{Ma_etal2021} also describe at a high level ongoing efforts towards similar goals on laser systems operating at repetition rate of a few Hz using a neural network approach.

There are ultra-intense laser systems that operate at higher than 1~Hz repetition rates, and these systems are already being used in efforts to accelerate protons. \citet{Morrison_etal2018}, for example, accelerated protons to $\sim$2~MeV energies using few mJ ultra-intense laser pulses at a repetition rate of 1~kHz. This experiment could potentially be replicated on the many mJ class, $\sim$kHz repetition rate laser systems that exist today (e.g. \citet{Cao2023}). It is also true that future industrial or defense applications of ultra-intense laser systems will likely operate closer to this repetition rate regime \citep{Palmer2018}. Ideally, we would like to train ML models on these systems in quasi-real time. A natural question is therefore, of the ML models that are being used today, can they be quickly trained on tens of thousands of shots or more? Can this be achieved with modest computational resources such as a single GPU, or would it require a supercomputer or GPU cluster? 
The ability to train a ML model in quasi-real time could greatly assist efforts to optimize and/or control the properties of ion beams resulting from the laser interaction. Specifically, the ML model could help discover ways to increase the max proton energy, or it could help with the inverse problem of wanting a particular ion energy distribution and needing to know the laser parameters to produce that distribution.

In our study, we will use three different ML methods to train models using up to 20000 synthetic data points with different levels of added noise. Although neural networks can approximate highly complex functions, it is not obvious that this method will be more accurate other methods because neural networks require large amounts of data to train. 

Our synthetic dataset is generated from a theoretical model that we modified from the theory model in \citet{Fuchs2005}. Although this synthetic data model is smoother and likely less featured than real experimental data sets, it allows us to generate a practically unlimited number of data points in order to test different ML models and examine how they perform with different assumptions about the noise.

Another approach, that some groups have investigated \citep{ liquidleaf,deeplearning, Dolier_etal2022, Djordjevic_etal2023},  would be to perform large numbers of 1D PIC simulations and to use this as training data for ML models. This is very different from our approach in that the noise comes from the finite number of particles in the PIC simulations whereas in our study the noise is added intentionally to represent what a real experiment might see. Another reason why we did not perform large numbers of 1D PIC simulations is that these simulations do not (without further assumption) capture the behavior of putting the target closer or further away from peak focus, as our model does. Placing the target off of peak focus is a potentially important strategy for controlling the proton energy spectrum (e.g. \citet{Morrison_etal2018,Loughran_etal2023}). 

In \S\ref{sec:synthetic}, we discuss our modified \citet{Fuchs2005} model and the noise added to it when generating the synthetic data set. In \S\ref{sec:ml}, we describe the three different machine learning models used in this study. In \S\ref{sec:results}, we show our results for the accuracy of the trained models. \S\ref{sec:perf} considers the computational performance and memory consumption of the trained models. In \S\ref{sec:opt} we use these trained models with a simple optimization task.
In \S\ref{sec:discuss} and \S\ref{sec:concl}, we summarize and conclude.

\section{Synthetic Data}
\label{sec:synthetic}

\subsection{Modified Fuchs et al. Model}

We generate synthetic data based on a physical model described by \citet{Fuchs2005} which introduces a cutoff time to the plasma expansion model by \citet{Mora_2003}. This model has five input parameters: laser intensity, wavelength, pulse duration, target thickness, and spot size. This model uses empirical formulae to estimate quantities such as laser absorption and hot electron temperature.
We extend the model (in a trivial way) by adding focal distance as an input and using the ideal Gaussian beam formula to determine the effective spot size and intensity. Had we used large numbers of 1D or even 2D PIC simulations instead of a theoretical model, a similar prescription would need to be added to model the diminished intensity of the laser pulse as the target moves away from peak focus.

As will be discussed later, in our study we kept the pulse duration and wavelength fixed. The spot size was also ``fixed" in the sense that the spot size at peak focus was always the same, but for non-zero focal distance the effective spot size on target depends on the distance from the target to peak focus. This approach of keeping the pulse duration and laser wavelength fixed while moving the target through the focus of the laser is similar to experimental studies like \citet{Morrison_etal2018} and \citet{Loughran_etal2023}, which both found interesting features in the proton acceleration results even when the target was not at the peak focus.

The \citet{Fuchs2005} model provides an estimate of both the maximum proton energy as well as the total proton energy and average proton energy. In this way, the Fuchs model provides three outputs. To be clear, these are the protons accelerated by the laser interaction, not necessarily the protons in the bulk target that do not gain significant energy.

There are many semi-analytic models that exist that could have been used for this purpose (e.g. \citet{Schreiber_etal2006,Passoni_Lontano2008,Passoni_etal2010,Zimmer_etal2021}) and that are potentially more accurate than \citet{Fuchs2005}, especially at high intensity. We used \citet{Fuchs2005} because it is well known in the field, it is relatively simple, and because we are restricting our domain of interest to protons accelerated by the Target Normal Sheath Acceleration (TNSA) mechanism (e.g. \citet{Clark_etal2000,Hatchett_etal2000,Snavely_etal2000,Passoni_etal2010}).

We did make one substantial modification of the Fuchs model to improve agreement with experiments. In order to improve the predicted maximum proton energy in the intensity regime between $10^{18}$~W~cm$^{-2}$ and $10^{19}$~W~cm$^{-2}$, we changed the relationship between the acceleration timescale of the protons ($\tau_{\rm acc}$) and the laser pulse duration ($\tau_{\rm laser}$).
\begin{equation}
    \tau_{\rm acc} = 4.0 \cdot \tau_{\rm laser}
\end{equation}
This modification is inspired by \citet{Djordjevic_etal2021}, who in their \S 5.3 treated the multiplier between the laser pulse duration and the proton acceleration timescale as a free parameter to be constrained by 1D Particle-in-Cell simulation results. Without this modification (i.e. the unmodified Fuchs et al. model), we found that the max proton energies in this intensity regime were too low compared to experiments in this intensity regime (e.g. \citet{Morrison_etal2018}).

Note that in computing the total proton energy and average proton energy from the \citet{Fuchs2005} model, we assumed the minimum kinetic energy to be zero for simplicity rather than using a non-zero cutoff.

\subsection{Range of Synthetic Data Generated}

We generate a synthetic data based on a laser system that can generate 40 femtosecond pulses, a spot size of 1.8 microns Full-Width Half-Max, and target thicknesses between 0.5 microns and 10 microns.  These parameters are similar to that of \citet{Morrison_etal2018}. We generated synthetic data with total pulse energies between 1.41~mJ and 14.14~mJ. At peak focus, these pulses correspond to intensities of $10^{18}$~W~cm$^{-2}$ and $10^{19}$~W~cm$^{-2}$ respectively. The focal distance was varied between -10 microns and +10 microns. Consequently the intensity on target for the generated points ranges from $\sim10^{17}$~W~cm$^{-2}$ to $10^{19}$~W~cm$^{-2}$. 

We generated 25000 data points by randomly sampling this parameter space.
The highest proton energy in the data set was near 2.3~MeV while the lowest proton energy was 1.6~keV. The random sampling of the parameter space is an important detail because a real, kHz repetition rate laser experiment would sample the parameter space in a very specific way, for example, by scanning through the focal distance or translating the target to monotonically investigate different thicknesses.  We intentionally sampled the parameter space in a randomized way to avoid ``gaps" in the synthetic data set that might bias the results. The random sampling of the intensity was exponentiated uniform, so as not to focus too much on high intensity or low intensity, while the random sampling of other parameters was uniform across the range. In future work, we will investigate more realistic synthetic data sets that better mimic how data from real kHz laser experiments are gathered. 

We use up to 20000 of the generated data points for training the ML models and reserve the remaining 5000 points for testing. Our decision to use 20000 data points as the maximum number of points in the training set, rather than set a higher or lower number, was informed by a few different priorities. On a 1~Hz repetition rate laser system like the one described in \citet{Loughran_etal2023}, collecting 20000 data points would take 5.5 hours, assuming continuous operation. Currently, there are a number of high-power laser systems operating near 1~Hz repetition rate, so 20000 data points represent roughly one full day of experimental time on these systems. 

As mentioned earlier, we are also interested in kHz repetition rate systems (e.g. \citet{Morrison_etal2018}). These systems produce 20000 shots in 20 seconds and 3.6 Million shots in an hour, but a number of limitations ranging from data acquisition systems to diagnostics to changing the laser parameters or the limited speed that the target can be moved can mean that consecutive shots on a kHz system probe essentially the same conditions. These and other practical limitations can easily cause an hour long experiment on a kHz laser system to only produce tens of thousands of shots with distinctly different conditions.

Focusing on up to 20000 points is therefore relevant to both repetition rate regimes and it is a stepping stone towards the  million-plus shot data sets that kHz laser systems with upgraded data acquisition and diagnostics will eventually produce.

\subsection{Noise model}
\label{sec:noise}

To better represent experimental data, we added noise to all three output quantities of the model -- max proton energy, total proton energy, and average proton energy. 
Our noise model involved making Fuchs model predictions for our parameter space and sampling from a log-normal distribution using \emph{each} prediction as the mean. We generated data sets with different noise levels by assuming that the standard deviation is between 5\% and 30\% of the mean value. When, in later sections, we refer to a 10\% Gaussian noise level, for example, this means that we generated data by sampling a log-normal distribution assuming that the standard deviation was 10\% of the mean. The ``Gaussian" in this context refers to the Gaussian function that underlies the log-normal distribution.  Because the standard deviation is assumed to be some fraction of the mean, as the Fuchs model predictions become larger, the noise level also becomes proportionally larger.

We include with this publication the the Python code that was used to generate the synthetic data and the code that was used to train the models \citep{desai_2024_10912447}.

\section{Machine Learning Methods}
\label{sec:ml}

We use the synthetic data to train three different machine learning models. These are Neural Network (NN), Support Vector Regression (SVR), and Gaussian Process Regression (GPR). In contrast to a neural network, SVR and GPR apply a ``kernel" function to map data to a higher dimensional space.

There are a number of caveats and important details in comparing these models, and we do not attempt to make a definitive comparison. What we do aim to do is to make reasonable efforts to compare the methods and their performance on a single GPU. Specifically, we use a NVIDIA Volta V100 GPU on the Pitzer cluster at the Ohio Supercomputer Center with 32~GB of GPU memory. Unless otherwise noted, we present results from training the ML models in single precision. 

As mentioned earlier, there are three independent variables in the data set -- target thickness, intensity and focal distance. First, we applied a logarithm to both the intensity and the three outputs of the model: maximum proton energy, total proton energy, average proton energy. Then, we applied standard z-score normalization \citep{han2011data} to all three inputs of the training set. We also apply z-score scaling to the outputs using the standard deviation and the mean of the training set. We note that the log-scaling of the outputs introduces a bias that causes under-prediction \citep{miller_1984}. In this paper, we multiply the detransformed outputs by a correction factor equal to the mean of the training data outputs divided by the detransformed outputs.

It bears mentioning that we did not ``pre-train" any of the ML models prior to training these models on the synthetic data described in \S\ref{sec:synthetic}.

\subsection{Support Vector Regression}

SVR is based on Support Vector Machines which were developed for classification problems. In SVR, one specifies a tolerance value that is used to determine which points follow the relation given by a particular curve and which points are considered outliers. This makes the task of finding the best fit curve like a classification problem. For more information about SVR, see \citet{Smola2004ATO}. 

To accelerate the calculations with GPU, we used the cuML library \citep{cuml}, which is part of the RAPIDS package. The version for our cuML library was 22.10.01. Our investigation used an epsilon of 0.01, tolerance of 0.001, and regularization parameter C of 2.5. Appendix~\ref{ap:hidden} describes results that justify these choices of hyperparameters. Additionally, we used the radial basis function (RBF) kernel. The $\gamma$ value is set by cuML's ``scale" option, which automatically sets the $\gamma$ value equal to 1 / (number of features * input variance).  For more details of our specific implementation please see the attached Python file \citep{desai_2024_10912447}. 

\subsection{Neural Network}
\label{sec:nn}

\begin{figure}
    \centering
    \includegraphics[width=4in]{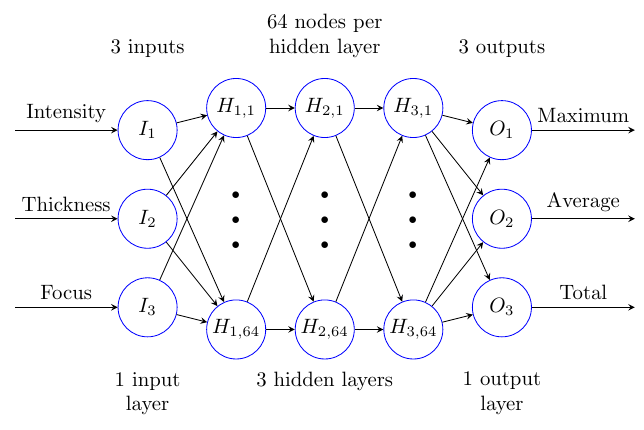}
    \caption{Architecture of the neural network model. The three inputs (intensity, thickness, and focal distance) and three outputs (maximum, total, and average proton energy) are fully connected by three hidden layers of 64 neurons each.}
    \label{fig:nn_arch}
\end{figure}

We used PyTorch \citep{NEURIPS2019_bdbca288} version 1.12.1 to implement a neural network (NN) model using a three hidden layer architecture described in Fig.~\ref{fig:nn_arch} with 64 neurons per layer. Appendix~\ref{ap:hidden} describes results that justify our choice of using three hidden layers.
We used a ``leaky'' ReLU activation function between fully connected nodes and trained in a batch size of 256. 

We train the neural network using the ``Adam" scheme for back-propagation as described in \citet{kingma2017adam}. We use an initial learning rate of 0.001, which the Adam scheme can increase or decrease as the model improves. 20 percent of the training set is used as a validation set, so that training can automatically stop when the validation loss stops decreasing.

\subsection{Gaussian Process Regression}
\label{sec:gp}

Gaussian Process Regression (GPR) is a Bayesian method which differs substantially from SVR and NN. The approach works by selecting a set of analytic functions that go through a series of data points. The GPR model can make a data-informed predictions by taking the weighted average of these functions, with the weights being the likelihood that any one of these functions is the ``true" function that correctly captures the model. A potential advantage over SVR and NN is that the GPR model uncertainty can be simply determined by the variance of these functions without any additional work. \citet{Schulz2017ATO} provides an excellent introduction to GPR.

To accelerate the computations with GPU, we used the GPyTorch library \citep{gardner2018gpytorch} version 1.9.1. We found this to be faster than the GPFlow library for our problem case.  As discussed in \citet{gardner2018gpytorch}, numerical instabilities can arise in cases with low or zero noise which can affect the accuracy of the GPR model. To avoid this we configured GPyTorch to use double precision floating point data and the other models were trained with double precision floating point data to be consistent. We used the RBF kernel for our GPR model with 30 training iterations and a learning rate of 0.2.

\section{Accuracy of Trained Models}
\label{sec:results}

\begin{figure*}
    \centering
    \includegraphics[width=4.0in]{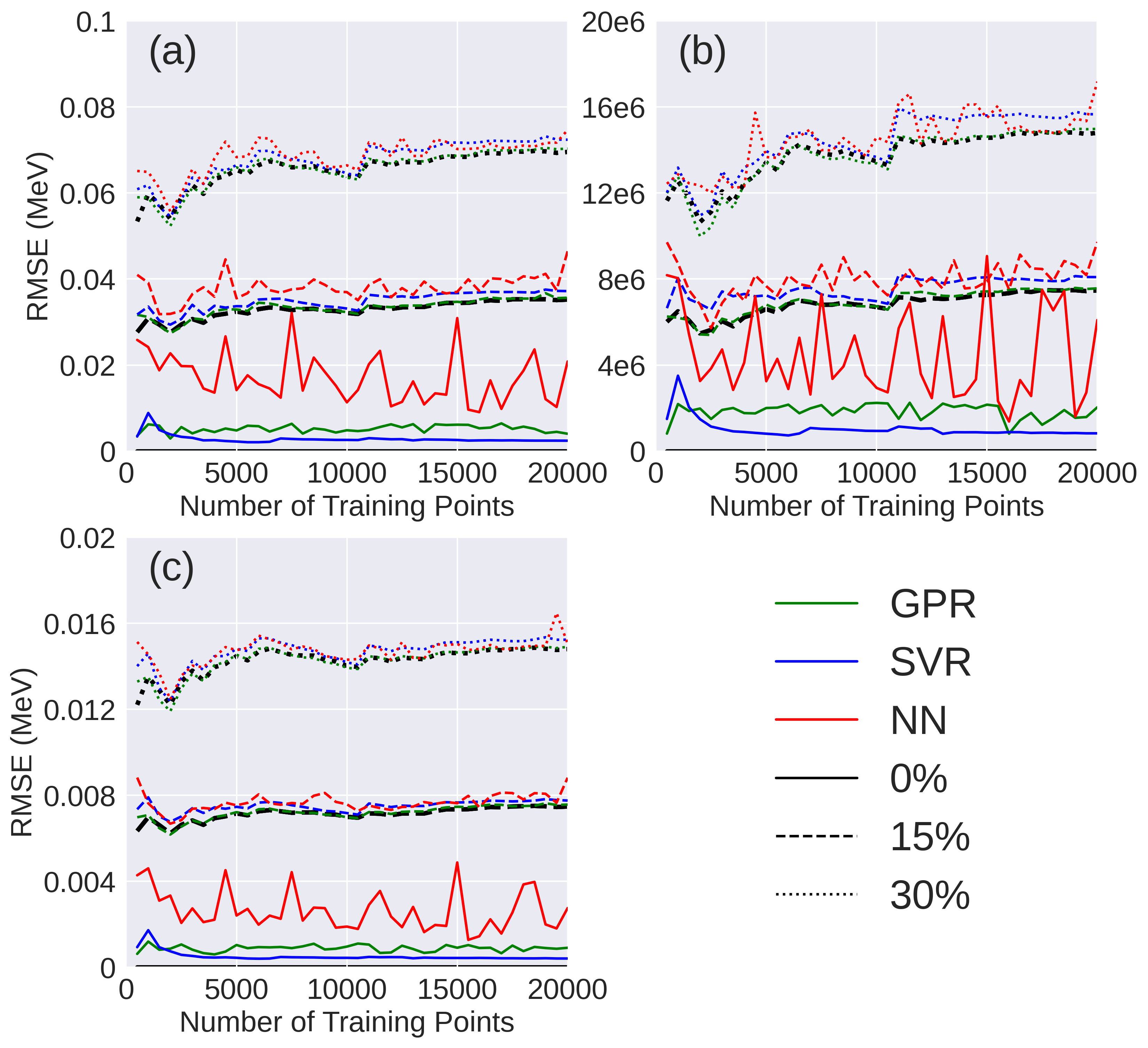}
    \vspace{-0.3cm}
    \caption{RMS error versus number of training points between ML model predictions for (a) max proton energy, (b) total proton energy, (c) average proton energy and noisy testing data. Each panel shows results from (solid) 0\%, (dashed) 15\% and (dotted) 30\% added noise in the data. Black lines with different line types indicate the RMS error between the noisy and noiseless data. Because we only compare ML models to noisy data in this figure, these black lines indicate the best that any ML model could conceivably do.}
    \label{fig:rms_accuracy}
\end{figure*}

Fig.~\ref{fig:rms_accuracy} highlights our results for the accuracy of the three different ML models in terms of the root mean square error (RMSE) for predicting the max proton energy (left panel), total proton energy (center panel) and average proto energy (right panel). Each panel shows results for each ML method from three different levels of Gaussian noise.

This plot was made by comparing trained ML models to up to 5000 testing points that were not included in the training data. Black lines with different line types show the RMSE between noisy and noiseless data. We include this to indicate the best that any ML model could conceivably do when compared to noisy data. Each line type corresponds to a different noise level.

\begin{figure*}
    \centering
    \includegraphics[width=4.0in]{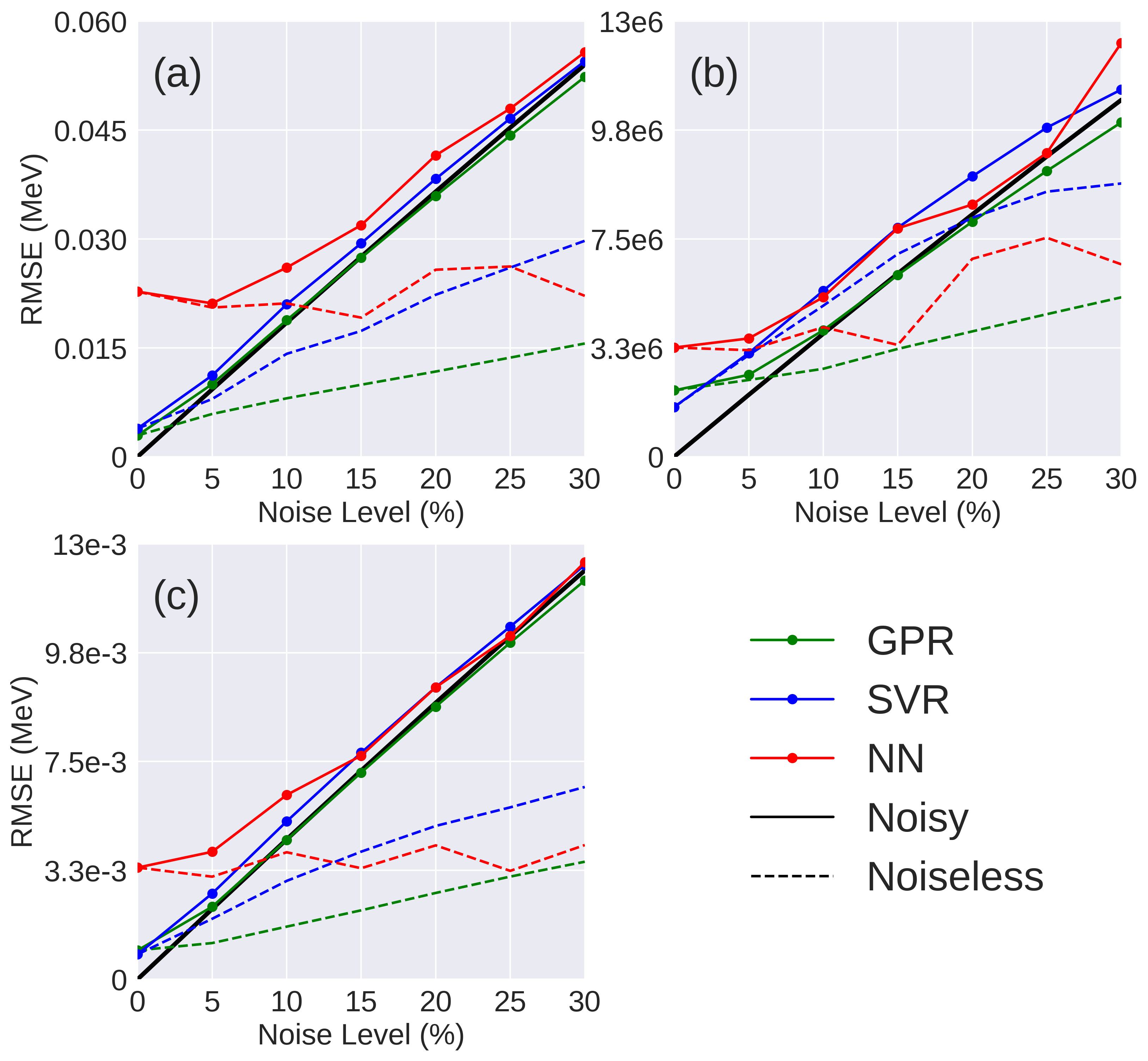}
    \caption{Solid lines show the typical RMS error in (a) max proton energy, (b) total proton energy, and (c) average proton energy when the ML models (which were trained on 2000 synthetic data points with noise) are compared to data with different levels of noise. Dashed lines show the typical error when those same ML models are compared to noiseless test data. Black solid lines indicate the RMS error between the noisy and noiseless data.}
    \label{fig:err}
\end{figure*}

We are interested in determining not only whether the ML models are reasonably accurate compared to the known level of noise in our synthetic data set, but also how well the ML models (which are trained on noisy data) compare to noiseless data. Fig.~\ref{fig:err} compares the accuracy of the ML models trained on 2000 data points with between 0\% to 30\% Gaussian noise compared to 500 test data points that both do and do not include noise. For comparison, a diagonal dotted black line shows the RMSE between noisy and noiseless data as a function of the noise level. This is helpful to interpret the comparison between the ML models and the noisy data as a kind of best case scenario, and in the comparison to noiseless data, to the extent that the ML models fall below this diagnonal dotted black line it indicates that the ML models are successfully averaging out the noise.

The figure shows that when the ML models are compared to the noisy data, except for the low noise NN results, the RMS error tends to be dominated by the noise as expected. Compared to noiseless data, the ML models (which were trained on noisy data) tend to have an RMS error that is smaller than the data sets they were trained on. Again, the only exception is the NN results for $\leq$5\% Gaussian noise.

\section{Performance and Memory Consumption}
\label{sec:perf}

\begin{figure}
    \centering
    \includegraphics[width=3.5in]{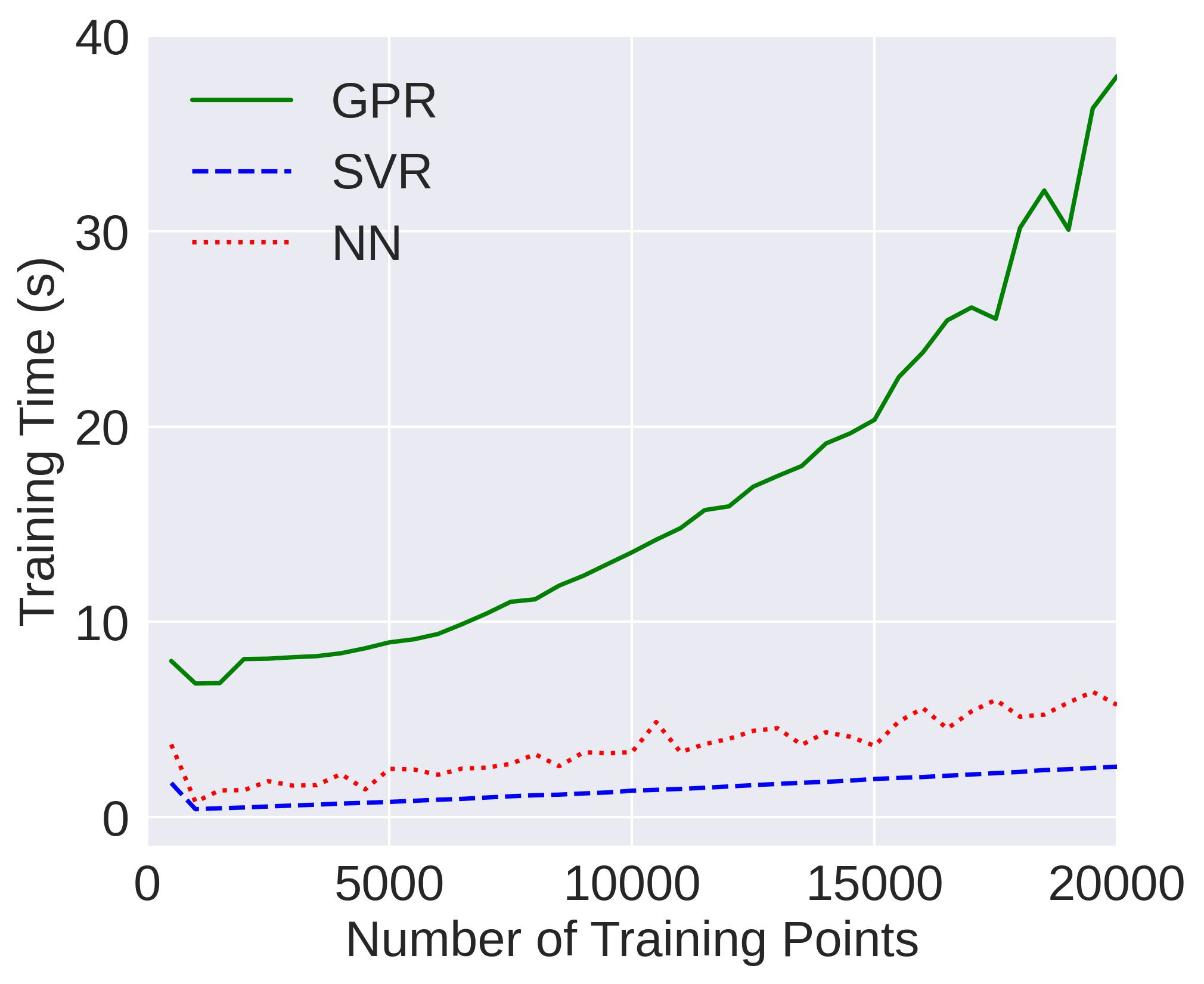}
    \caption{Comparing the execution time of the different ML models averaged across noise levels in computing the maximum, total, and average proton energies.}
    \label{fig:performance}
\end{figure}

Fig.~\ref{fig:performance} shows our primary results for the execution time of the different ML models using between 500 and 20000 synthetic training data points averaged between noise levels of 0 to 30\%.

Of the three models, the clear winner with respect to execution time is SVR, followed by NN and, last, GPR. Although we expected the execution time for GPR to approach $O(N^3)$ \citep{wang2022intuitive}, the increase in the execution time with increasing numbers of training data points was not as steep as this, at least for the range of $N$ that we probed and the data set we used. 

Table~\ref{tab:memory} summarizes the average GPU memory consumption of the three different ML models on 20000 synthetic data points averaged across noise levels. The NN had the lowest memory consumption while GPR consumed the most memory.

\begin{table}
    \centering
    \begin{tabular}{c} 
    \end {tabular}
    \begin{tabular}{cccc}
         & SVR & NN & GPR    \\
         \hline
       GPU Memory Utilization (GiB)  & 1.6 & 1.1  & 14.0    \\
       \hline
    \end{tabular}
    \caption{Average GPU memory consumption results when training on data with 20000 points.
    }
    \label{tab:memory}
\end{table}

\section{Optimization Task}
\label{sec:opt}

In this section we use trained models to perform a simple optimization task which is to determine a set of input parameters that would produce a proton spectrum up to a particular cutoff energy -- and that would provide the highest possible numbers of protons in that energy range despite the limited amount of laser energy available. This is a simple but interesting task because one can produce a proton energy spectrum in a particular energy range by placing the target at peak focus and decreasing the laser energy and intensity, or one can keep the laser energy fixed and move the target out of focus until the intensity produces the desired proton energy range, or one can keep the target at peak focus and increase the thickness until the proton energy range decreases to the desired range. So although our modified Fuchs et al. function is relatively simple this optimization task should produce a non-trivial result. 

We perform this optimization by re-casting it as a minimization problem. Producing a proton energy spectrum with a max kinetic energy of $KE_{\rm cutoff,goal}$ and with a significant laser energy to proton energy conversion efficiency is like minimizing this function

\begin{eqnarray}
     f(KE_{\rm cutoff},\eta_{proton}) & =  &   \frac{|KE_{\rm cutoff} - KE_{\rm cutoff,goal}|}{KE_{\rm cutoff, goal}} \nonumber  \\
     & & + \frac{C}{\eta_{proton}} + g( KE_{\rm cutoff},KE_{\rm cutoff,goal}) \nonumber \\ \label{eq:function}
\end{eqnarray}
where $C$ is a parameter that is introduced to ensure that both the first and second terms have similar importance. $g(KE_{\rm cutoff},KE_{\rm cutoff,goal})$ is a penalty term that is set to a large positive number when $KE_\text{cutoff}$ is more than $\pm$15$\%$ away from $KE_\text{cutoff, goal}$. 

The ML models provide estimates of $KE_{\rm cutoff}$ and $\eta_{proton}$ given the input parameters of laser energy, distance from peak focus to the target and target thickness. We therefore vary these three parameters and use the ML models to determine where in this 3D parameter space the function defined by Eq.~\ref{eq:function} is a minimum. We then compare to performing this same task for the modified Fuchs et al. model without any noise to obtain the true values of the parameters that minimize the function. The goal of this exercise is to determine if the ML methods, which demonstrate different levels of accuracy as discussed in \S\ref{sec:results}, are more or less robust to noise in performing this task. 

\begin{figure}
    \centering
    \includegraphics[width=4in]{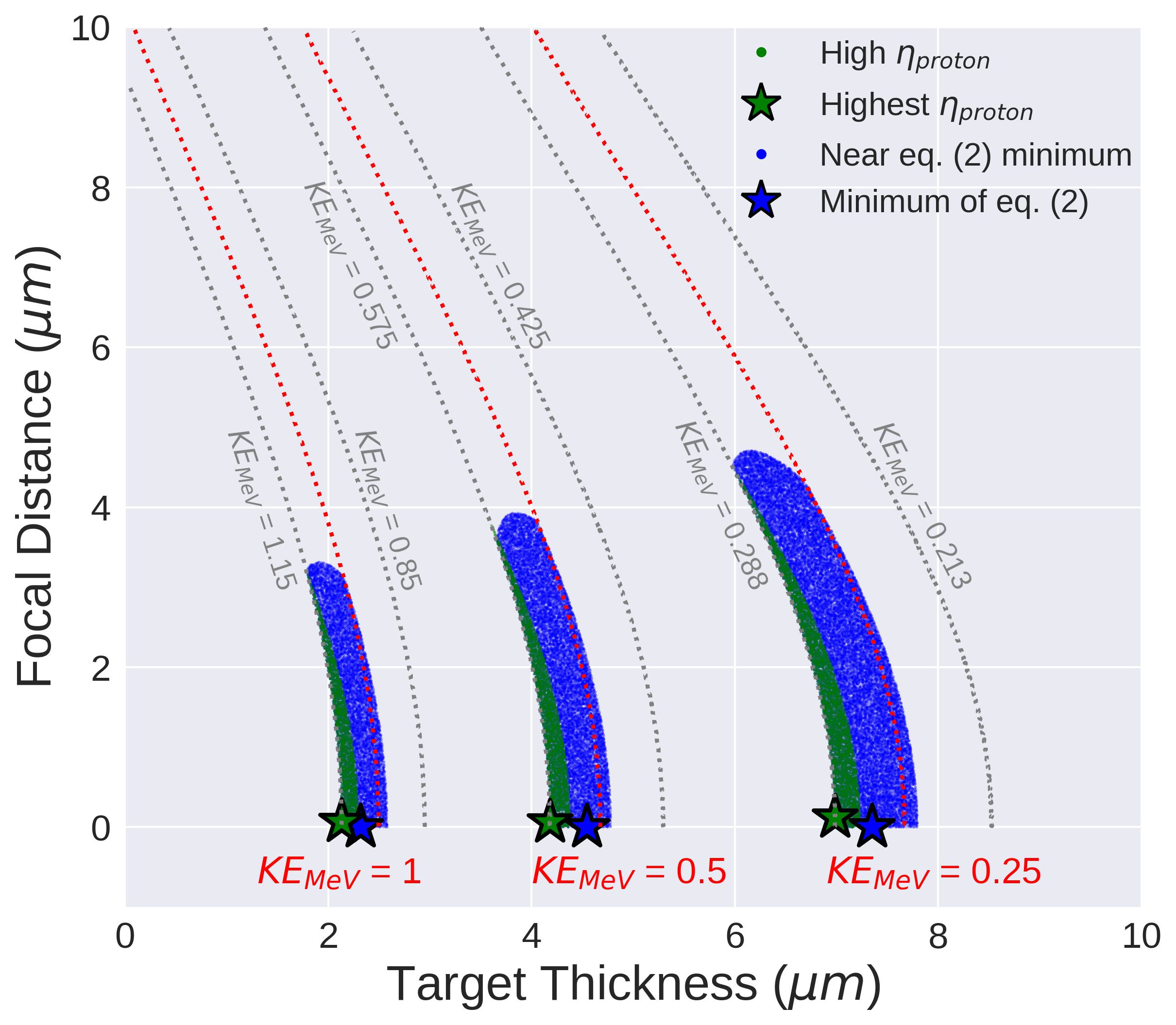}
    \caption{Parameters that produce maximum proton energy cutoffs in three different desired ranges: 1.0~MeV, 0.5~MeV and 0.25~MeV. Combinations of thickness and focal distance that produce these energy cutoffs (irrespective of the laser to proton conversion efficiency) are shown with dotted red lines. With each red line we also show with dotted gray lines the thicknesses and focal distances that produce proton energy cutoffs that are +15\% or -15\% of the cutoff goal. Green shaded areas show regions where the laser to proton conversion efficiency is high (i.e. within 5\% of the optimal value). A green star shows the ideal conditions for maximizing the proton conversion efficiency. The blue region corresponds to using all the terms in Eq.~\ref{eq:function} and the blue star indicates the ideal conditions according to that minimization scheme.}
    \label{fig:banana}
\end{figure}

To the extent that the ML models perform this task well their conclusions regarding the ideal conditions for three different maximum kinetic energy cutoffs should resemble the results shown in Fig.~\ref{fig:banana}, which is not from any trained ML model but is derived from the analytic model itself without any noise. The three different maximum kinetic energy cutoffs are 0.25~MeV, 0.5~MeV and 1.0~MeV. 

\begin{figure*}
    \centering
    \includegraphics[width=5in]{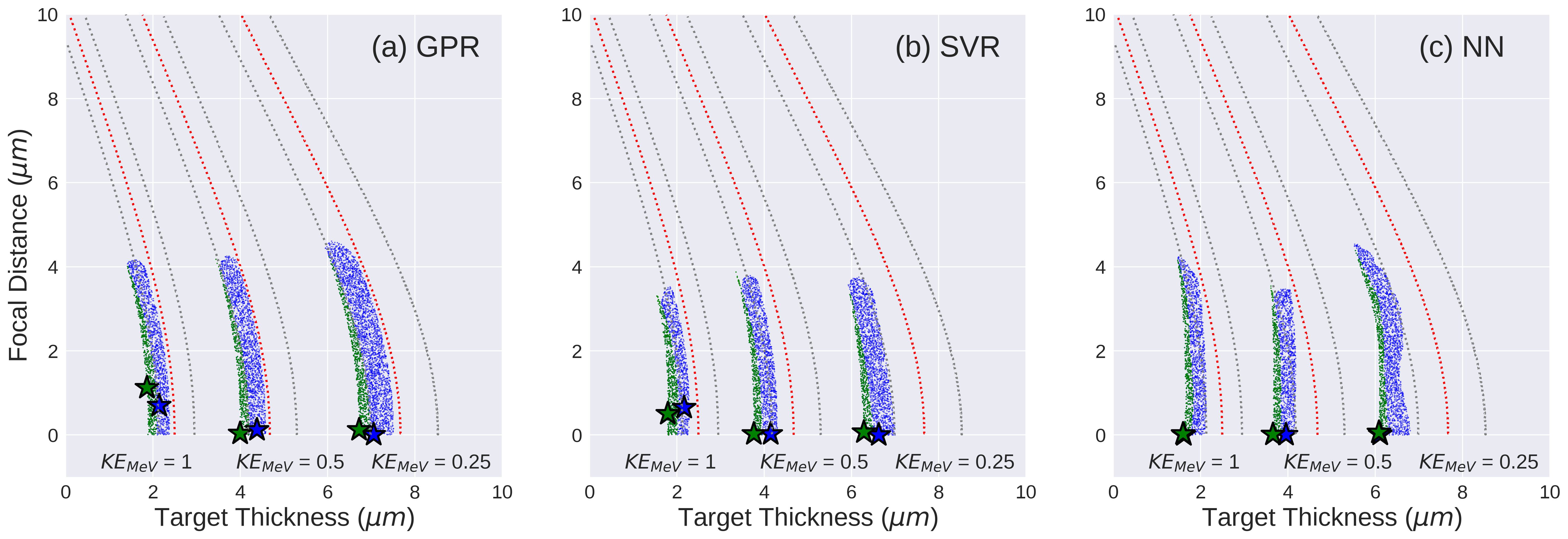}
    \caption{Parameters that produce maximum proton energy cutoffs according to three trained ML Models on 2000 data points with 30 $\%$ added noise: (a) GPR, (b) SVR, (c) NN compared against the red and gray lines plotted in Fig.\ref{fig:banana}. The green and blue shaded regions are scatter plots of a subset of evaluated points that fall within 5 $\%$ of the model's predicted optimum according to the same criteria in Fig. \ref{fig:banana}.}
    \label{fig:banana_ml}
\end{figure*}

There are many different combinations of target thickness and focal distance that produce proton spectra with these cutoffs and the thin red lines in Fig.~\ref{fig:banana} indicate those possibilities for each cutoff energy. In some applications, having a kinetic energy cutoff that is 15\% higher or lower than the target cutoff may be acceptable, so we also indicate on the plot with gray lines the thicknesses and focal distances that produce energies that are 15\% above or below the three desired cutoff energies.

Although there are many combinations of target thickness and focal distance that can produce a proton energy spectrum with the right cutoff, not all of these combinations also optimize the laser to proton energy conversion efficiency ($\eta_{proton}$). Because $\eta_{proton}$ tends to increase for increasing maximum kinetic energy cutoffs, in Eq.~\ref{eq:function} we chose $C = 0.8 \cdot 10^{-3}, 1.8 \cdot 10^{-3} \text{ and } 4.5 \cdot 10^{-3}$ respectively for cutoff energies 0.25~MeV, 0.5~MeV and 1.0~MeV to ensure that the optimization prioritizes both matching the maximum kinetic energy cutoff and increasing $\eta_{proton}$. 
The green areas in Fig.~\ref{fig:banana} shows the part of the parameter space that would achieve a proton energy cutoff within $\pm$15\% of the cutoff goal \emph{and} have a significant $\eta_{proton}$ value. Generally, in our analytic model, the need to optimize $\eta_{proton}$ favors focal distances near zero and thinner targets. A green star shows the ideal target thickness and focal distance given these constraints. This is equivalent to using Eq.~\ref{eq:function} as the optimization function but without the first term. The results from using all the terms in Eq.~\ref{eq:function} is shown with a blue shaded area. Specifically this area includes the minimum of Eq.~\ref{eq:function} as well as conditions that cause the optimization function to be within 5\% of the global minimum.
Although similar to the green area, now the optimization will favor conditions that have both a significant $\eta_{proton}$ and which produce cutoffs closer to desired energy. The ideal thickness and focal distance including all the terms in Eq.~\ref{eq:function} is shown with a blue star. Generally the blue star is close to the green star except that the blue star is always closer to the desired cutoff energy (red line). 

Because the analytic function we are using yields the same results for positive and negative focal distances, we only consider positive focal distances. In a real experiment one would need to explore both positive and negative focal distances, and the contours that optimize proton energy generation will likely have more structure because of pre-pulse effects that are especially important at peak focus (zero focal distance).

Whereas Fig.~\ref{fig:banana} shows results for optimum conditions from our analytic model without any noise, Fig. \ref{fig:banana_ml} shows results from the same exercise using the three ML models. To help with the comparison the three sets of lines in Fig.~\ref{fig:banana} also appear in Fig.~\ref{fig:banana_ml}. Specifically we show results from training the three models on synthetic data with 30\% added noise and with 2000 points in the training set. Analogously to Fig.~\ref{fig:banana}, we can repeatedly query these trained ML models to obtain their predictions. We then use these predictions with Eq.~\ref{eq:function} and obtain both the optimum values and the regions that fall within 5\% of that optimum value.

Fig.~\ref{fig:banana_ml} shows that both SVR and NN tend to find optima that occur at thinner target configurations than the gray lines would indicate. This means that these ML models incorrectly favor conditions that would produce protons with more than 15\% of the intended cutoff energy. More generally, Fig.~\ref{fig:banana_ml} shows that there are practical consequences for using ML models that perform better or worse on the tests described in \S\ref{sec:results}.

\section{Discussion}
\label{sec:discuss}

Of the three ML models, the neural network tended to be less accurate than the other models (Figs.~\ref{fig:rms_accuracy}, \ref{fig:err} \& ~\ref{fig:banana_ml}). As a comment on this result, it is important to note that the approach of the NN method is very different from either SVR and GPR in that it is essentially trying to fit 8771 free parameters using 20000 data points or less. Even with sophisticated methods to determine those parameters like ``Adam" backpropagation \citep{kingma2017adam}, this is not enough to reach high accuracy without more data, and we did not do any ``pre-training" because this would have required additional assumptions and potentially introduce bias. A reasonable question is whether our choice of using 3 hidden layers (Fig.~\ref{fig:nn_arch}) could have affected this result. In Appendix~\ref{ap:hidden} we answer this question by conducting a grid search through different hidden layers and neurons to optimize the network architecture.

We examined the performance of the ML models running on one GPU. We found that the SVR training time was much less than GPR and NN, which is perhaps not surprising since SVR has a reputation for being a fast method (e.g. \citet{towardsdatascience,Xu_etal2011}). The performance results made us optimistic that at least some ML models can be accurately trained using one GPU in quasi-real time in kHz ultra-intense laser experiments.

We found that the GPR method consumed the most GPU RAM of the three models. Efforts to use GPR in ultra-intense laser experiments that produce significantly more than 20000 data points on a short timescale may need to worry about GPU RAM constraints, or instead use an approximate GPR method (e.g. \citet{svgp}).

To demonstrate how the quality of the ML model can lead to practical consequences and to provide a simple application of our framework, we used the trained ML models to predict the laser parameters that produce a desired proton energy spectrum while also maximizing  the laser to proton energy conversion efficiency. We performed such tests in Sec.~\ref{sec:opt} using 2000 training points and found (Fig.~\ref{fig:banana_ml}) that the NN and SVR models incorrectly favor thinner targets than would be needed to produce protons in a desired energy range and in some cases NN and SVR incorrectly favored a non-zero focal distance even though the analytic model (which does not have any pre-pulse physics) indicates that zero focal distance was always most favorable for maximizing the conversion efficiency.

All these results are based on a modified \citet{Fuchs2005} model that we developed to generate synthetic data. Overall, this model is relatively smooth and well behaved whereas real experiments typically have more features (e.g. \citet{Morrison_etal2018,Loughran_etal2023}). Certainly this limits the generality of our results. Our goal was to create a kind of standard candle for evaluating ML methods for TNSA experiments. In a forthcoming paper we are exploring larger and more realistic data sets with more physics (and features) included in the synthetic data model.

\section{Conclusions}
\label{sec:concl}

We tested three different machine learning models on a synthetic data set for laser-accelerated protons by training these models on up to 20000 synthetic data points. The data set was generated using a modified \citet{Fuchs2005} model that included Gaussian noise to simulate the kind of noise that could be present in a real experiment. The machine learning models were Gaussian Process Regression (GPR), Support Vector Regression (SVR) and a three-hidden-layer neural network (NN). 

Of the three methods, the NN models were the least accurate and this was especially true when the number of training points was low. This included an exercise where we used the ML models to predict optimum conditions for proton acceleration at different cutoff energies. The poor accuracy of the NN model likely stems from the large number of free parameters that must be precisely determined for the model to become highly accurate. We did not ``pre-train" any of the models to try to ameliorate this issue.

In terms of performance on one GPU, SVR was by far the fastest model for execution time. GPR was the slowest, taking 30 seconds to train on 20000 data points. The neural network used the least amount of GPU memory while GPR consumed 5-10 times more GPU memory than the other models. Generally the performance results suggest that quasi-real time training of these models on kHz repetition rate laser systems should be feasible, especially with SVR.

We stress that these results should not be regarded as the final word on the usefulness of these ML models in the context of laser acceleration of protons. More complex data sets may yield different results. Moreover, there are certainly ML models that deserve attention that we did not study here. 

We provide Jupyter notebooks with our Python code. We also provide the 25000 point synthetic data sets that we generated\citep{desai_2024_10912447}. By providing these files we hope to encourage others to compare other ML models against our results as a benchmark.

\section*{Acknowledgements}
Supercomputer allocations for this project included time from the Ohio Supercomputer Center. We also thank Pedro Gaxiola for help with early work with SVR. We acknowledge support provided by the National Science Foundation
(NSF) under Grant No. 2109222. Any opinions, findings, and conclusions or recommendations expressed in this material are those of the author(s) and do not necessarily reflect the views of the National Science Foundation. CO
and RO were supported in summer 2022 by the Air Force Office of Science Research summer faculty program. This work was supported by Air Force Office of Scientific Research (AFOSR) Award (PM: Dr. Andrew B. Stickrath). This
work was also supported by Department of Energy (PM: Dr. Kramer Akli).

\appendix

\clearpage
\section{On Choosing Optimal Hyper-parameters}
\label{ap:hidden}

The number of hidden layers is an important concern when designing a neural network. Work done by \citet{HORNIK1989359, LESHNO1993861} showed that neural networks with at least one hidden layer and using a non-polynomial activation function can serve as an universal approximator for any function given a sufficient number of nodes. Adding more hidden layers can, in some cases, reduce the training time and allow for faster convergence for a neural network. However, this depends on the data set with more complex data sets benefiting from additional hidden layers.

\begin{figure}
    \centering
    \includegraphics[width=4in]{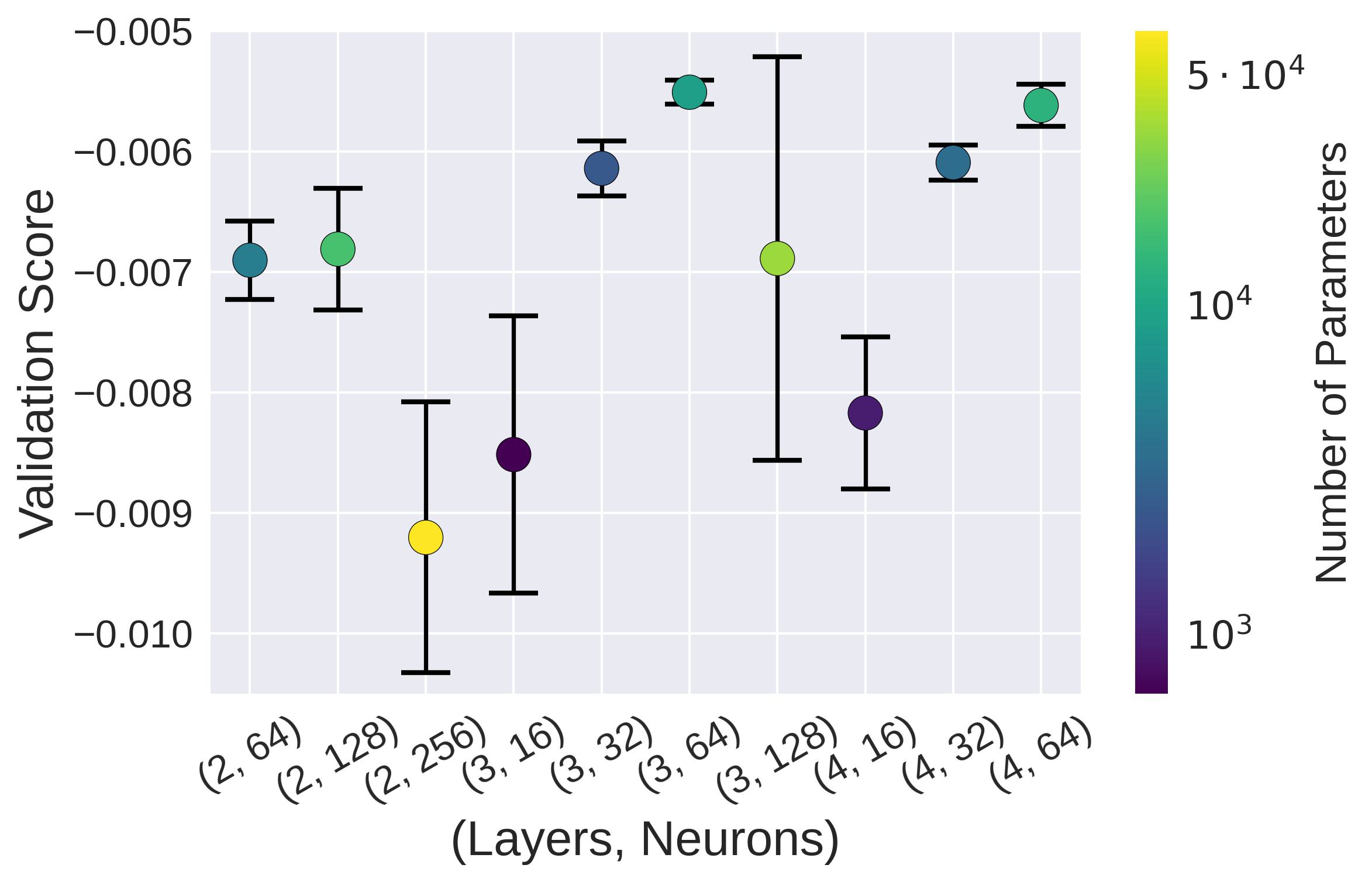}
    \caption{Validation score (using negative mean square error) with error bars from 5-fold cross-validation plotted against the number of hidden layers and neurons per layer from a grid search of hyper-parameters on a 5000 point synthetic dataset with 10\% added noise. The following hyper-parameters are fixed: batch size (256), learning rate decay $\gamma$ (0.98), activation function (LeakyReLU), optimizer (Adam). The coloring shows the number of learnable parameters (computed from Eq. \ref{eq:learnable_parameters} in the neural network model.}
    \label{fig:multi_layer}
\end{figure}

Our neural network model architecture was chosen by performing a grid search of model hyper parameters using scikit-learn's \citep{scikit-learn} \texttt{GridSearchCV} module. In Fig. \ref{fig:multi_layer}, various model architectures between 1 - 4 hidden layers and 16-256 neurons per layer were tested. The architectures with highest validation score have 3 or 4 hidden layers. Notably, the 4 hidden layer scores are not higher than the 3 hidden layer scores, demonstrating that adding more layers doesn't always improve the network.  The results with only 1 hidden layer all had an order of magnitude lower score than the scores shown and were not included as a result.

Additionally, the number of learnable parameters in the neural network model is shown in the color bar of Fig. \ref{fig:multi_layer}. This model has a learnable parameter for every connection between nodes in adjacent layers and additional bias parameters for every neuron in the output and hidden layers. This results in a total of

\begin{equation}
    f(n, \ell) = 7n + 3 + n(n+1)(\ell - 1) \label{eq:learnable_parameters}
\end{equation}

parameters where $n$ is the number of neurons per layer and $\ell$ is the total number of hidden layers. In Fig \ref{fig:multi_layer}, we can see that more learnable parameters does not necessarily equate to a better model. 

\begin{figure}
    \centering
    \includegraphics[width=3.5in]{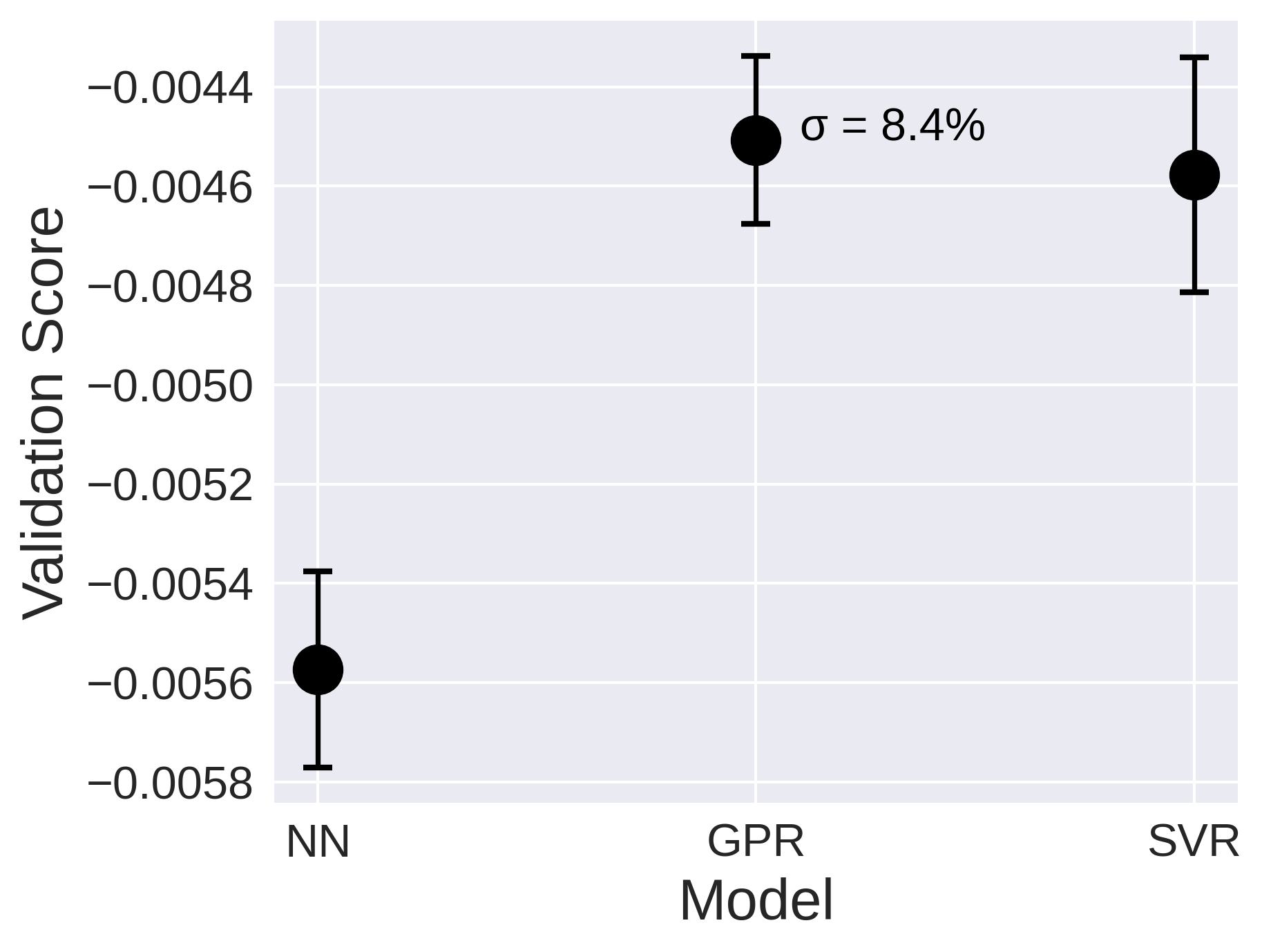}
    \caption{Validation score comparison with error bars from 5-fold cross-validation for the SVR, NN, and GPR models from a grid search of hyper-parameters on a 5000 point synthetic dataset with 10\% added noise. }
    \label{fig:grid_search}
\end{figure}

For the SVR and GPR, we also used \texttt{GridSearchCV} to settle on the optimal hyper-parameters chosen in Section \ref{sec:ml} and the results can be seen in Fig. \ref{fig:grid_search}. 

Here, the GPR narrowly outperforms the SVR which both strongly outperform the NN model. An added benefit of the GPR model is its inherent uncertainty estimation given by $\sigma = 8.4\%$ (can be compared to the noise level $10 \%$ of the training data) which was determined by averaging the uncertainty estimate over all points in the training dataset. When data are expensive to obtain, the GPR's uncertainty estimate can guide one's choice of new data points to take (where the uncertainty is greater).

\bibliographystyle{jpp}

\bibliography{main}

\end{document}